\newcommand{\beq}{\begin{quote}}
\newcommand{\enq}{\end{quote}}
\newcommand{\be}{\begin{equation}}
\newcommand{\en}{\end{equation}}
\newcommand{\del}{\delta}

\newcommand{\eps}{\epsilon}
\newcommand {\la}{\lambda}

\documentclass[11pt]{amsart}
\usepackage{geometry}                % See geometry.pdf to learn the layout options. There are lots.
\usepackage{graphicx}
\usepackage{amssymb}
\usepackage{epstopdf}
\title{ A simple derivation  of the two force laws for elliptic orbits from Proposition 6 in  Newton's  {\it  Principia}  }
\author{Michael Nauenberg}
\begin{document}
\begin{abstract} 
Based on Propostion 6 of his {\it Principia}, Newton's  geometrical derivation in  Propositions 10 and 11 for  the radial dependence of the
two central forces that lead to  elliptical orbits is notoriously difficult. An alternate and more transparent derivation 
is obtained  by applying  the affine transformation of a circle into an ellipse.

\end{abstract}
\maketitle

%\section{Introduction}
\section{Introduction}

In 1684, when Edmond Halley visited Isaac Newton  and asked him if the knew the orbital curve for an inverse square force, Newton promptly responded that it is an ellipse. But when asked for his calculation,  he could not find it, and when he tried  the derivation again
he could  not reproduce it \cite{westfall}.  But  several months later he sent him a short treatise entitled  {\it De Motu Corporum Gyratum}
containing  a proof, which  appeared  in his {\it Principia} as Proposition 11. His proof, based on a general expression
for central forces in Proposition 6,  is  notoriously difficult to understand.
For example,  in the``Guide to Newton {\it Principia}", I. B. Cohen devoted  six pages to describe it
 \cite{cohen}, while  in the ``The Key to Newton's Dynamics",  Bruce Brackenridge took  eleven pages for the same task \cite{bruce}. Even  Richard Feynman complained that he couldn't follow Newton's proof  \cite{feynman},   and developed an alternate one  that turned out to have been given  previously by J. C Maxwell \cite{maxwell}  who in turn attributed it to  Sir William Hamilton.  At  the start of $18th$ century, however,  able mathematicians like Jacob Hermann, Pierre  Varignon and Johann Bernoulli  were able to express Newton's relation for central force in Proposition 6 as a differential equation for the orbit  in  Cartesian coordinates, while  Gotfried Leibniz  obtained such an  equation  in polar coordinates \cite{michael}.

Given a continuous   curve and a fixed point, in Proposition  6  Newton  gave an expression in geometrical form for the attractive central force $f $ acting on  a body that is  confined to move along this curve,  describing  areas proportional to the time elapsed
according to Proposition 1, and in Propositions 10 and 11 he treated the cases when this curve is an  ellipse.
In Proposition 10 the center of force is placed at the center of the ellipse,  and he proved that the resulting force depends linearly on the distance from this center, while in Proposition 11 he treated the case relevant to planetary motion when the center of force is at a focus  of the ellipse, and proved that the force varies inversely
with the square of the distance from the center. 

 Referring to the diagram for Proposition 6  reproduced in Fig.1,  Newton's  relation for a central force is
\be
\label{force1}
f\propto \frac{QR}{SP^2 \times QT^2},
\en
where SP is the radial distance of a body at P, revolving around the center of force at S on the orbital curve APQ,
 Q is  near P,  and QT is a line perpendicular to SP. The location of R appears to be at  the intersection of  the  line ZPY tangent to the curve at  P  with the extension of the radial line SQ, and then  QR=SR-SQ.   
  The product $SP\times QT$ is the  area of the triangle SPQ,  which  is
proportional to the time interval for the motion of a body  from P to Q, according to Proposition 1. When
 the  curve APQ is expressed  in polar coordinates ($r,\theta$),  it is  straightforward to calculate this  differential area:
\be
SP\times QT= r(\theta)^2\del \theta.
\en
where $\del \theta$ is the first order  differential angle between the radial lines $SP=r(\theta)$ and $SQ=r(\theta +\del \theta )$, (see Fig. 2).
The difficult problem is to evaluate QR which must be  a second order differential  for the central force 
$f$, Eq.\ref {force1},  to exist in the continuum limit  when $\del \theta \rightarrow 0$.
 In the first edition  of the {\it Principia} (1687),   Proposition 6
  states: ``QR should be drawn parallel to the distance SP" \cite{bruce},  as  shown also  in the diagrams associated with
  Propositions 10 and 11.  But  the diagram  in Fig.1 associated with Proposition 6,  shows QR drawn parallel to SQ.  This difference
  does not have any practical  consequence,  because QP is  a first order  differential that becomes arbitrarily small in the continuum limit,
   but  it  matters for  the method adopted to evaluate the magnitude of QR. In Propositions 10 and 11, Newton took QR parallel to SP.
   
In Section 2  the affine transformation  that relates the ellipse to a circle is applied to obtain the displacement   QR , and the radial dependence of the force f, Eq. \ref{force1}, is obtained when  the center of force is located at a focus of the ellipse corresponding
to Proposition 11.
In Section 3 the  case corresponding to Proposition 10 is treated when the center of force is at the center of the ellipse,  
corresponding to Proposition 10.  An Appendix contains  a brief historical account of  Robert Hooke's remarkable graphic and analytic treatment of this problem in 1685.

\clearpage

  \begin{figure}[htbp] %  figure placement: here, top, bottom, or page
   \centering
   \includegraphics[width=6in]{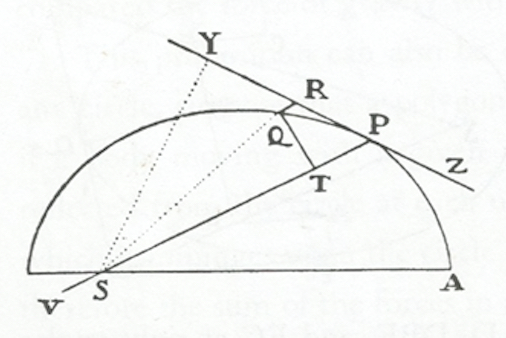} 
   \caption{ Diagram for Proposition 6
   }
\end{figure}

%\section{Inverse Square force}
\section{Proposition 11, Inverse Square force}
Referring to Fig.2,  $x_P'$, $y_P'$  are  the Cartesian coordinates  of a point P' on the circle of radius $a=CA$
centered at C, and  $x_P=x_P'-a\epsilon$ and $y_P=\lambda y_P'$ are  the corresponding coordinates of a point  P on the ellipse
obtained by the affine transformation with parameter $\lambda$.  The ellipse coordinates are given
relative to the focus of the ellipse at S, where $CS=a\eps$,  $\la =\sqrt{1-\eps^2}$, and $\eps$ 
 is the eccentricity of the ellipse.  Then
\be
\label{tan}
tan \theta=\frac{y_P}{x_P}= \frac{\lambda sin\theta'}{(cos\theta' - \epsilon)}
\en
gives the relation between the  angles $\theta$ and $\theta'$ for the radial lines r=SP and r'=CP' relative to CA. We have
$r=a(1-\eps cos\theta')$, and substituting $ cos\theta'=(cos \theta+\eps)/(1+\eps cos\theta)$ one obtains
\be
\label{polar}
r=\frac{a(1-\eps^2)}{1+\eps cos\theta},
\en
the well known equation for the ellipse in polar coordinates.

The first order differential angle $\del \theta$  between the nearby radial lines SP and SQ,  and $\del \theta'$ between the lines CP' and CQ'  are related by
\be
\frac {\del \theta}{\cos^2\theta}=\frac{\lambda \delta \theta'(1-\epsilon cos\theta')}{(cos \theta' -\epsilon)^2)}.
\en

Substituting $ cos\theta' =(r/a) cos \theta+\epsilon$, and $r/a=(1-\epsilon^2)/(1+\epsilon cos \theta$) for the equation of
the  ellipse in polar coordinates,  Eq. \ref{polar}, one finds
\be
\label{delth}
\del \theta=  \frac{\la a}{r}\del \theta'
\en

Newton's measure for the central force f in Proposition  6 is 
\be
\label{force}
f \propto \frac{QR}{QT^2 \times SP^2},
\en
where $QT\times SP= r^2 \del \theta $ is  the differential area, and according to Eq.\ref {delth}
\be
\label{diff}
r^2\del \theta=r b \del \theta',
\en
where $b=\la a$ is the minor axis of the ellipse.

The point R is located at the  intersection of the tangent line at P and the extension of the radial line CQ; hence
 QR=SR-SQ (see Fig.1).  Since a  tangent line on the circle is orthogonal to the radial line,  it is straightforward to locate the intersection  
R' of  the tangent line at P' with the extension of the line CQ', and we have
\be
CR'=\sqrt{a^2+(a \del \theta')^2} = a(1+\frac{\del \theta'^2}{2})
\en
to second order in $\del \theta'$.
Hence
\be
\label{qr} 
Q'R'=CR'-CQ'= \frac{1}{2}a\del \theta'^2 , 
\en
and by the affine transformation
 \be
 \label{qr}
 QR=\frac{1}{2} a\del \theta'^2 \sqrt{cos^2\theta' +\la^2 sin^2 \theta'},
 \en
where
\be
cos \theta' = \frac{(\eps+cos\theta)}{(1+\eps cos\theta)},
\en
according to Eq. \ref{tan}.
Substituting this relation between the cosines of the angles $\theta$ and $\theta'$ in Eq.\ref{qr}, one finds that
\be
QR=Q'R'=\frac{1}{2} a \del \theta'^2
\en
The diagram in Fig.2  shows that QR  does not lay along the extension of SR and differs in magnitude from Q'R',
but this relation  is valid because $\del \theta $ and $\del \theta' $  are first order differentials.
Hence, applying  Newton's expression for force, Eq. \ref{force}, and  for the differential area, Eq. \ref{diff} ,  we obtain

\be
f =\frac{QR}{SP^2\times QT^2}=\frac {a}{2r^2b^2} =\frac{1}{SP^2\times L},
\en
in accordance with Proposition 11, where $L=2 BC^2/AC$ is the {\it latus rectum} of the ellipse.

  \begin{figure}[htbp] %  figure placement: here, top, bottom, or page
   \centering
   \includegraphics[width=6in]{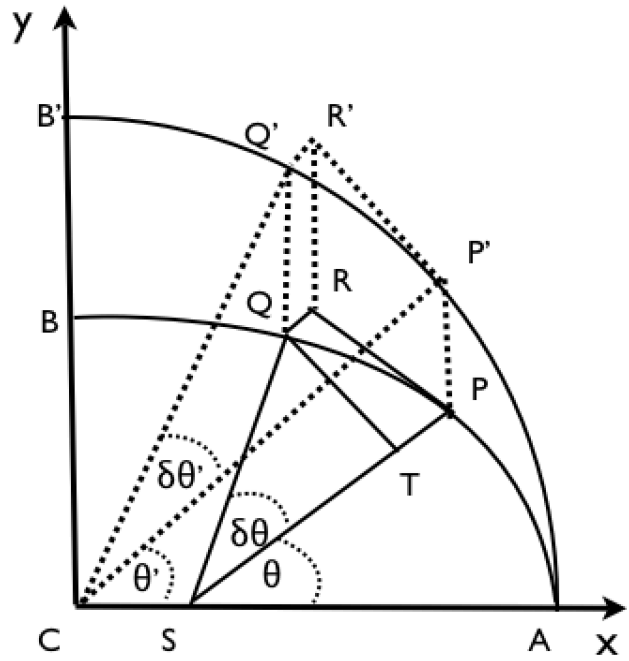} 
   \caption{ AP'Q'B' is a quarter of a circle with center at C and radius CA,  APQB is the corresponding section
   of an ellipse obtained  from the affine transformation  x=x', y=$\lambda y$, where 
   $\la=\sqrt{1-\eps^2}$, and  S is a focus of the ellipse  where CS=$\eps CA$.
   }
\end{figure}
\clearpage

%\section{Linear force}
\section{ Proposition 10, Linear force}

Referring to Fig.3, $x_P'=acos\theta'$, $y_P'=asin\theta'$ are the Cartesian coordinates  of   P' on the circle of radius $a$, and by the
affine transformation,  $x_P=x_P', y_P=\lambda y_P'$  are the  corresponding coordinates  of P on the ellipse with major axis $a= CA$, 
 and minor axis $b=\la a=BC$.
  Then
\be
\frac{y_P}{x_P} = tan \theta= \la tan \theta',
\en
and
\be
\frac {\del \theta}{\cos^2\theta}=\frac{\lambda \delta \theta'}{cos^2 \theta'}
\en
 The radial distance $r=PC$ is 
\be
r=a\sqrt{cos^2 \theta'+\la^2 sin^2 \theta'}=\frac{a}{\sqrt{cos^2 \theta+sin^2\theta/\lambda^2}}
\en
and  
\be
QR=Q'R' \sqrt{cos^2 \theta'+\la^2 sin^2 \theta'},
\en
where according to Eq.\ref{qr} $Q'R'=a \del \theta'^2/2$. Therefore
\be
QR=\frac{\la r\del \theta'^2}{2}.  
\en
and the differential area
\be
QT \times PC= r^2 \del \theta=a^2 \la \del \theta'.
\en
Hence, the central force
\be
f= \frac{QR}{QT^2\times PC^2}=\frac{r}{2b^2 a^2}=\frac{PC}{2BC^2AC^2}
\en
in accordance with Newton's result in Proposition 10.

  \begin{figure}[htbp] %  figure placement: here, top, bottom, or page
   \centering
   \includegraphics[width=4in]{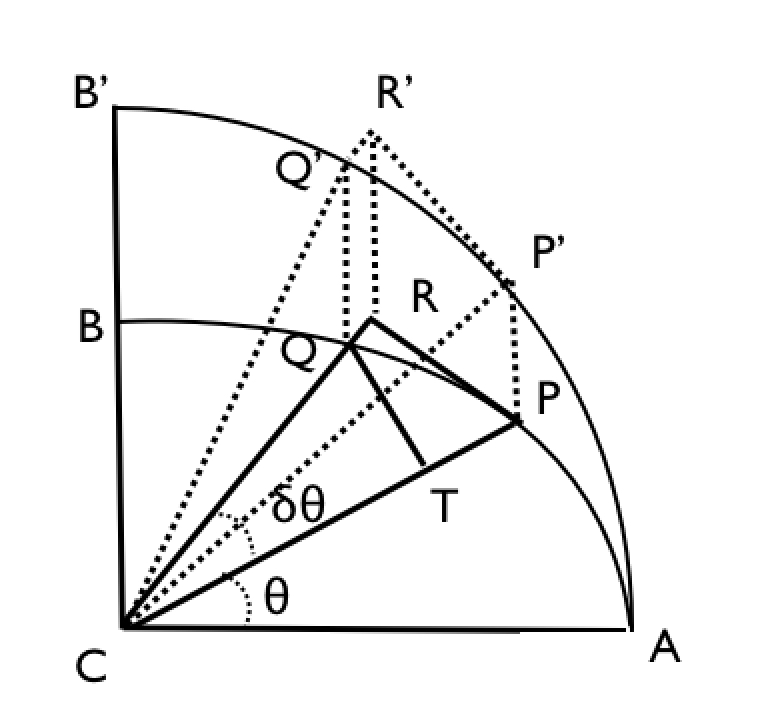} 
   \caption{ AP'Q'B' is a quarter of a circle with center at C and radius CA,  APQB is the corresponding section
   of an ellipse obtained  from the affine transformation  x=x', y=$\lambda y$
 }
\end{figure}
\clearpage

\section*{Appendix}

 In 1685, when Newton sent the  first draft of his {\it Principia} entitled {\it De Motu Corporum Gyrum} to the Royal Society,  Robert Hooke who was  its secretary  at the time,  recognized that Newton's geometrical construction in Theorem 1  could be applied  as a graphical method to calculate orbits\cite{hooke1}, \cite{michael5}.  He  then  proceeded to draw such an  orbit for  a periodic sequence
 of impacts  that depended linearly on the distance from the center of force, and obtained a discrete polygon with the vertices located on an ellipse\cite{hooke2}.   Six years earlier, he had communicated
 to Newton his own ideas about the  nature of gravitational forces that accounted for planetary motion along the lines
 that Newton afterwards  implemented.  Part of Hooke's proof that the  vertices of the discrete orbit  the had obtained were located on an ellipse  was based on the affine transformation of a circle into an ellipse (see Figs. 2 and 3  in reference \cite {hooke1}). But his proof  differs from Newton's proof in Proposition 10 which assumes first the orbital curve  to be an ellipse, and  then determines the radial dependence of the force that gives rise to such an  orbit when the center of force is located at the center of the ellipse. Hooke concluded the discussion associated with his diagram with the remark that  `` the polygone becomes various according to the differing degrees of Gravity at Different distances from the center". It is therefore likely that he would have attempted to obtain graphically also the orbit for an inverse square force - the case of interest for gravitation. But there isn't any evidence among the manuscripts of Hooke that have been preserved that he carried out this calculation. If he had tried to carry  out this calculation graphically
 with similar initial conditions,   he would have found that only the resulting vertices of the discrete orbit for the
 first seven impacts are located on an ellipse .  Afterwards,  I have shown that  the graph for this  orbit diverges which would have presented a puzzle for  Hooke  \cite{hooke3}. Perhaps for this reason he did not publish his remarkable graphic results.


\begin{thebibliography}{30}
\bibitem{hooke2}   Among Hooke's papers  in the Trinity Library, Cambridge, there is a manuscript dated Sept. 1685  that
contains Hooke's drawing of a discrete  polygonal orbit with vertices on an ellipse, reproduced in Ref. \cite{hooke1}.
\end{thebibliography}
\end{document}